\documentclass[aps,prd,draft,showpacs]{revtex4}

\begin{document}

\title{Renormalization Group Evolution of the Effective Potential in the Broken
Symmetry Phase}
\author{Chungku Kim}
\date{\today}

\begin{abstract}
We investigate the renormalization group(RG) evolution for the neutral
scalar field theory in the broken symmetry phase. By using the minimum
condition of the vacuum expectation value(VEV), we show that the RG evlution
of the effective potential in the broken symmetry phase is governed by the
same RG functions in case of the symmetric phase.
\end{abstract}
\maketitle

\affiliation{Department of Physics, Keimyung University, Daegu 704-701, KOREA%
} \setcounter{page}{1}

The effective potential plays an important role in studies of the vacuum
instability, dynamical symmetry breaking, and the dynamics of composite
particles\cite{Sher}. The scalar field of a quantum field theory with a
spontaneously broken symmetry have a non-vanishing VEV which provide a mass
to the Higgs particle, gauge bosons and fermions. The effective potential in
the broken symmetry phase is expressed in term of the Higgs particle which
have a vanishing VEV. Recently, it was shown that the effective potential in
the broken symmetry phase is scale invariant in the case of vanishing Higgs
field which implies the scale invariance of the physical cosmological
constant\cite{Foot}. In this paper, we will investigate the RG evolution of
effective potential in the broken symmetry phase in case of non-vanishing
Higgs field. Although we will consider the neutral scalar field theory for
simplicity, the generalization to more complicated cases will be
straightforward.

The effective potential of the neutral scalar field theory in the minimal
subtraction scheme(MS) is independent of the renormalization mass scale $\mu 
$ and satisfies the RG equation\cite{RG} 
\begin{equation}
(D+\gamma ^{MS}\phi \frac{\partial }{\partial \phi })V_{MS}(\mu ,\lambda
,m^{2},\phi )=0.
\end{equation}
where 
\begin{equation}
D\equiv \mu \frac{\partial }{\partial \mu }+\beta _{\lambda }^{MS}\frac{%
\partial }{\partial \lambda }+\beta _{m^{2}}^{MS}\frac{\partial }{\partial
m^{2}}
\end{equation}
In the case $m^{2}<0$, the scalar field $\phi $ have a non-vanishing VEV $v$
satisfying $_{{}}$%
\begin{equation}
\left[ \frac{\partial V_{MS}(\mu ,\lambda ,m^{2},\phi )}{\partial \phi }%
\right] _{\phi =v}=0,
\end{equation}
from which one can determine $v$ as a function of $\mu ,\lambda $ and $m^{2}.
$ Then the Higgs field $\sigma $ with a vanishing VEV is defined as $\sigma
\equiv \phi -v$ and the effective potential in the broken symmetry phase $%
V_{SSB}(\mu ,\lambda ,m^{2},\sigma )$ is given by 
\begin{equation}
V_{SSB}(\mu ,\lambda ,m^{2},\sigma )\equiv V_{MS}(\mu ,\lambda ,m^{2},\sigma
+v(\mu ,\lambda ,m^{2}))
\end{equation}
Recently, it was shown that when $\sigma =0,$ $V_{SSB}(\mu ,\lambda ,m^{2},0)
$ satisfies 
\begin{equation}
DV_{SSB}(\mu ,\lambda ,m^{2},0)=0
\end{equation}
\cite{Foot} which implies the scale invariance of the physical cosmological
constant. In order to investigate the RG evolution of $V_{SSB}(\mu ,\lambda
,m^{2},\sigma )$ in the case of non-vanishing Higgs field $(\sigma \neq 0)$,
it is necessarily to obtain the RG evolution of the $v(\mu ,\lambda ,m^{2})$%
. Actually, the RG evolution of the VEV plays an important role in
understanding the upper bound of the Higgs boson mass\cite{bound} and the
CKM matrix\cite{CKM} and since the one-loop calculation\cite{One}, the
perturbative calculations showed that the VEV of a scalar field have a same
RG evolution as the corresponding Higgs field\cite{zwan}. In order to obtain
the RG evolution of the VEV of a scalar field from the minimum condition
given in eq.(3), let us take a derivative of eq.(1) with respect to $\phi $
to obtain 
\begin{equation}
(D+\gamma ^{MS}\phi \frac{\partial }{\partial \phi }+\gamma ^{MS})\frac{%
\partial V_{MS}(\mu ,\lambda ,m^{2},\phi )}{\partial \phi }=0.
\end{equation}
By substituting $\phi =v(\mu ,\lambda ,m^{2})$ in this equation and by using
eq.(3), we obtain 
\begin{equation}
\left[ (D+\gamma ^{MS}\phi \frac{\partial }{\partial \phi })\frac{\partial
V_{MS}(\mu ,\lambda ,m^{2},\phi )}{\partial \phi }\right] _{\phi =v(\mu
,\lambda ,m^{2})}=0
\end{equation}
and by applying the operation $D$ to eq.(3), we obtain an
\begin{equation}
D\left[ \frac{\partial V_{MS}(\mu ,\lambda ,m^{2},\phi )}{\partial \phi }%
\right] _{\phi =v(\mu ,\lambda ,m^{2})}=0
\end{equation}
Note that in eq.(7), the operator $D$ is inside of the bracket $\left[
...\right] _{\phi =v(\mu ,\lambda ,m^{2})}$ and hence does not act on $v(\mu
,\lambda ,m^{2})$ whereas in eq.(8), it is outside of the bracket $\left[
...\right] _{\phi =v(\mu ,\lambda ,m^{2})}$ and act on $v(\mu ,\lambda
,m^{2})$. Then, we can write eq.(8) as 
\begin{equation}
\left[ D\frac{\partial V_{MS}(\mu ,\lambda ,m^{2},\phi )}{\partial \phi }%
\right] _{\phi =v(\mu ,\lambda ,m^{2})}+\frac{\partial ^{2}V_{MS}(\mu
,\lambda ,m^{2},v)}{\partial v^{2}}Dv=0
\end{equation}
By substituting eq.(7) to first term of eq.(9), we obtain 
\begin{equation}
\left[ \frac{\partial ^{2}V_{MS}(\mu ,\lambda ,m^{2},\phi )}{\partial \phi
^{2}}\right] _{\phi =v(\mu ,\lambda ,m^{2})}(Dv-\gamma ^{MS}v)=0
\end{equation}
Since $\left[ \frac{\partial ^{2}V_{MS}(\mu ,\lambda ,m^{2},\phi )}{\partial
\phi ^{2}}\right] _{\phi =v(\mu ,\lambda ,m^{2})}$ is an arbitrary quantity,
we conclude that the RG evolution of the $v(\mu ,\lambda ,m^{2})$ is given
by 
\begin{equation}
Dv(\mu ,\lambda ,m^{2})=\gamma ^{MS}v(\mu ,\lambda ,m^{2})
\end{equation}
Usually, this result was expected from the argument that in the broken
symmetry phase, VEV is renormalized by the wave function renormalization
constant of the Higgs field irrespective of the functional form of $v(\mu
,\lambda ,m^{2})$. But it is clear that eq.(11) cannot be satisfied by
arbitrary function $v(\mu ,\lambda ,m^{2})$.

Now, in order to obtain the RG evolution of the effective potential in the
broken symmetry phase, let us apply $D+\gamma ^{MS}\sigma \frac{\partial }{%
\partial \sigma }$ to $V_{SSB}(\mu ,\lambda ,m^{2},\sigma )$ : 
\begin{eqnarray}
&&(D+\gamma ^{MS}\sigma \frac{\partial }{\partial \sigma })V_{SSB}(\mu
,\lambda ,m^{2},\sigma )=\left[ DV_{MS}(\mu ,\lambda ,m^{2},\phi )\right]
_{\phi =\sigma +v(\mu ,\lambda ,m^{2})}  \nonumber \\
&&+\left[ \frac{\partial V_{MS}(\mu ,\lambda ,m^{2},\phi )}{\partial \phi }%
\right] _{\phi =\sigma +v(\mu ,\lambda ,m^{2})}Dv+\gamma ^{MS}\sigma \frac{%
\partial }{\partial \sigma }\left[ V_{MS}(\mu ,\lambda ,m^{2},\phi )\right]
_{\phi =\sigma +v(\mu ,\lambda ,m^{2})}
\end{eqnarray}
By using eq.(11), we can combine the last two terms of the above equation as 
\begin{equation}
\gamma ^{MS}v\left[ \frac{\partial V_{MS}(\mu ,\lambda ,m^{2},\phi )}{%
\partial \phi }\right] _{\phi =\sigma +v(\mu ,\lambda ,m^{2})}+\gamma
^{MS}\sigma \frac{\partial }{\partial \sigma }\left[ V_{MS}(\mu ,\lambda
,m^{2},\phi )\right] _{\phi =\sigma +v(\mu ,\lambda ,m^{2})}=\left[ \gamma
^{MS}\phi \frac{\partial V_{MS}(\mu ,\lambda ,m^{2},\phi )}{\partial \phi }%
\right] _{\phi =\sigma +v(\mu ,\lambda ,m^{2})}
\end{equation}
Then, by substituting this result into eq.(12) and by using eq.(1), we
obtain 
\begin{equation}
(D+\gamma ^{MS}\sigma \frac{\partial }{\partial \sigma })V_{SSB}(\mu
,\lambda ,m^{2},\sigma )=\left[ (D+\gamma ^{MS}\phi \frac{\partial }{%
\partial \phi })V_{MS}(\mu ,\lambda ,m^{2},\phi )\right] _{\phi =\sigma
+v(\mu ,\lambda ,m^{2})}=0
\end{equation}
which means that the RG evolution of the effective potential in the broken
symmetry phase is governed by the same renormalization group functions in
the MS\ scheme. Actually, by using the two-loop effective action\cite{2loop}%
. 
\begin{eqnarray}
V_{MS}(\mu ,\lambda ,m^{2},\phi ) &=&\frac{1}{2}m^{2}\phi ^{2}+\frac{1}{24}%
\lambda \phi ^{4}+\Lambda +\frac{1}{4}\frac{\hbar }{(4\pi )^{2}}(m^{2}+\frac{%
1}{2}\lambda \phi ^{2})^{2}(L_{MS}-\frac{3}{2})  \nonumber \\
&&+\frac{\hbar ^{2}}{(4\pi )^{4}}\{(\frac{1}{8}\lambda m^{4}+\frac{1}{4}%
\lambda ^{2}m^{2}\phi ^{2}+\frac{3}{32}\lambda ^{3}\phi ^{4})L_{MS}^{2}+(-%
\frac{1}{4}\lambda m^{4}-\frac{3}{4}\lambda ^{2}m^{2}\phi ^{2}-\frac{5}{16}%
\lambda ^{3}\phi ^{4})L_{MS}  \nonumber \\
&&+(\frac{1}{8}\lambda m^{4}+(\frac{3}{4}+\Omega (1))\lambda ^{2}m^{2}\phi
^{2}+(\frac{11}{32}+\frac{\Omega (1)}{2})\lambda ^{3}\phi ^{4})\}+O(\hbar
^{3}),
\end{eqnarray}
where 
\begin{equation}
\Omega (1)=-\frac{1}{2\sqrt{3}}\sum_{n=1}^{\infty }\frac{1}{n^{2}}\sin (%
\frac{n\pi }{3})\simeq -0.293
\end{equation}
and 
\begin{equation}
L_{MS}\equiv \log \left( \frac{m^{2}+\frac{\lambda }{2}\phi ^{2}}{\mu ^{2}}%
\right) ,
\end{equation}
we obtain the VEV as 
\begin{equation}
v(\mu ,\lambda ,m^{2})=-m\sqrt{\frac{6}{\lambda }}\{1+\frac{1}{2}\frac{\hbar 
}{(4\pi )^{2}}\lambda (-L+1)+\frac{\hbar ^{2}}{(4\pi )^{4}}\lambda ^{2}(-%
\frac{1}{4}L^{2}+\frac{3}{2}L-5\Omega (1)-\frac{7}{4})\}+O(\hbar ^{3})
\end{equation}
where 
\begin{equation}
L\equiv \log \left( \frac{2m^{2}}{\mu ^{2}}\right) 
\end{equation}
Then, by using the RG functions given by\cite{RGfn} 
\begin{eqnarray}
\beta _{\lambda }^{MS} &=&\mu \frac{d\lambda }{d\mu }=3\frac{\hbar }{(4\pi
)^{2}}\lambda ^{2}-\frac{17}{3}\frac{\hbar ^{2}}{(4\pi )^{4}}\lambda ^{3}+(%
\frac{145}{8}+12\text{ }\varsigma (3))\frac{\hbar ^{3}}{(4\pi )^{6}}\lambda
^{4}+\cdot \cdot \cdot , \\
\beta _{m^{2}}^{MS} &=&\mu \frac{dm^{2}}{d\mu }=\frac{\hbar }{(4\pi )^{2}}%
\lambda -\frac{5}{6}\frac{\hbar ^{2}}{(4\pi )^{4}}\lambda ^{2}+\frac{7}{2}%
\frac{\hbar ^{3}}{(4\pi )^{6}}\lambda ^{3}+\cdot \cdot \cdot , \\
\beta _{\Lambda }^{MS} &=&\mu \frac{d\Lambda }{d\mu }=\frac{1}{2}\frac{\hbar 
}{(4\pi )^{2}}m^{4}+\frac{1}{16}\frac{\hbar ^{3}}{(4\pi )^{6}}%
l^{2}m^{4}\cdot \cdot \cdot ,
\end{eqnarray}
and

\begin{equation}
\gamma ^{MS}=\frac{\mu }{\phi }\frac{d\phi }{d\mu }=-\frac{1}{12}\frac{\hbar
^{2}}{(4\pi )^{4}}\lambda ^{2}+\frac{1}{16}\frac{\hbar ^{3}}{(4\pi )^{6}}%
\lambda ^{3}+\cdot \cdot \cdot .
\end{equation}
we can check that eq.(11) is satisfied and by substituting eq.(2) to eq.(4)
and by expanding in $\hbar $, we can see that eq.(14) is also satisfied up
to $O(\hbar ^{2})$. Actually, we have checked that eq.(14) is satisfied to $%
\hbar ^{3}$ order.

Finally, let us discuss the RG running of the Higgs mass and the coupling
constants in the broken symmetry phase. There are two typical schemes to
define the parameters in the broken symmetry phase containing $v$. One is
using only $v^{(0)}=-m\sqrt{\frac{6}{\lambda }}$ which is the tree level
value of VEV in the tree Lagrangian and the remaining terms $v-v^{(0)}$ act
as a finite counter-terms to remove the tadpole terms in the higher order
Feynman diagrams\cite{Taylor}. The other is using $v$ itself in the tree
Lagrangian and includes the tadpole diagrams in the higher order Feynman
diagrams\cite{Jegerlehner}. For example, in case of the coupling constant
for the cubic Higgs interaction $h\sigma ^{3}$ which were absent in the
symmetric phase, $h$ is defined by $-m\sqrt{\frac{\lambda }{6}}$ in the
former case and $\frac{1}{6}\lambda v$ in the latter case. Then, from
eqs.(11) and (14) we can obtain the RG functions as 
\begin{equation}
\beta _{h}=\mu \frac{\partial h}{\partial \mu }=(\frac{1}{2m}\beta
_{m^{2}}^{MS}+\frac{1}{2\lambda }\beta _{\lambda }^{MS})h
\end{equation}
in the former case and 
\begin{equation}
\beta _{h}=\mu \frac{\partial h}{\partial \mu }=(\gamma ^{MS}+\frac{1}{%
\lambda }\beta _{\lambda }^{MS})h
\end{equation}
in the latter case respectively. In case of the running Higgs mass term $%
\frac{1}{2}m_{H}^{2}\sigma ^{2}$, $m_{H}^{2}$ is defined by $2m^{2}$ in the
former case and $-m^{2}+\frac{\lambda }{2}v^{2}$ in the latter case. Then
the corresponding RG functions are given by 
\begin{equation}
\beta _{m_{H}^{2}}\equiv \mu \frac{\partial m_{H}^{2}}{\partial \mu }=2\beta
_{m^{2}}^{MS}
\end{equation}
in the former case and 
\begin{equation}
\beta _{m_{H}^{2}}=\mu \frac{\partial m_{H}^{2}}{\partial \mu }=(2\gamma
^{MS}+\frac{1}{\lambda }\beta _{\lambda }^{MS})\frac{\lambda }{2}v^{2}-\beta
_{m^{2}}^{MS}=(2\gamma ^{MS}+\frac{1}{\lambda }\beta _{\lambda
}^{MS})m_{H}^{2}+(2\gamma ^{MS}+\frac{1}{\lambda }\beta _{\lambda
}^{MS}-\gamma _{m^{2}}^{MS})m^{2}
\end{equation}
in the latter case respectively.

In summary, we have investigated the RG evolution of the VEV of a scalar
field from the minimum condition of the VEV and have shown that the RG
evolution of the effective potential of the spontaneously broken symmetry is
governed by the same renormalization group functions of a theory in case of
the symmetric phase. As a result, we can determine the RG functions of the
running Higgs mass and the coupling constant for the cubic Higgs interaction.
It is easy to show that the result of this paper can be 
extended to the theories which have two Higgs particles.

\end{document}